\documentclass[12pt]{article}
\usepackage{graphicx}
\usepackage{subcaption}
\usepackage{amsmath,amssymb,xcolor,physics}
\usepackage{url}
\usepackage{ulem}
\usepackage{hyperref}
\usepackage{tabularray}

\allowdisplaybreaks[4]

\begin{document}

\begin{center}

{\Large
Comments on ``Ether of Orbifolds"}

\end{center}

\vspace{0.2cm}
\begin{center}
Masanori Hanada
\end{center}

\vspace{0.3cm}
\begin{center}
School of Mathematical Sciences, Queen Mary University of London\\
Mile End Road, London, E1 4NS, United Kingdom
\end{center}

\vspace{0.5cm}

\begin{center}
  {\bf Abstract}
\end{center}
We comment on a recent manuscript ``Ether of Orbifolds" by Henry Lamm. In the first version, it was mistakenly claimed that the orbifold lattice Hamiltonian is not gauge invariant, and a quantity $\epsilon_g$, which has nothing to do with a non-existent ``gauge violation", was introduced. The scaling of this $\epsilon_g$ was used to claim a huge simulation cost. In fact, $\epsilon_g$ characterizes the shift of the effective lattice spacing --- because, in the orbifold lattice formulation, the lattice is generated dynamically from the vacuum expectation value of the complex matrices. In the second version, the claim about the gauge symmetry was partially corrected, based on our comments. However, $\epsilon_g$ is still mistakenly interpreted as a measure of ``departure from SU($N$)", inconsistently with the foundational results by Kaplan, Katz, and \"{U}nsal, and also by Arkani-Hamed, Cohen, and Georgi. This interpretation plays a central role in sustaining the argument introduced in the first version.

\newpage

In ref.~\cite{Lamm:2026zzl}, whose ver.~1 appeared on April 1$^{\textrm{st}}$  2026, Hank Lamm analyzed the efficiency (or inefficiency) of the orbifold lattice approach to quantum simulations~\cite{Buser:2020cvn,Bergner:2024qjl,Halimeh:2024bth,Hanada:2025yzx,Bergner:2025zkj,Halimeh:2025ivn,Hanada:2025goy}.\footnote{
The orbifold lattice Hamiltonian was introduced by Kaplan, Katz, and \"{U}nsal~\cite{Kaplan:2002wv} in the context of supersymmetry on a lattice. See refs.~\cite{Cohen:2003xe,Cohen:2003qw,Kaplan:2005ta} for related papers on spacetime lattices.  
} 
His analyses were based on multiple fundamental misconceptions. Among these, a crucial issue was the assertions regarding gauge symmetry. 
He introduced five terms (equations (1) to (5) in ref.~\cite{Lamm:2026zzl}), and claimed that (3), (4), and (5) ``enforce gauge symmetry". 
He also claimed (3) is ``a Gauss-law–like constraint enforcing tracelessness." Based on these elementary errors, he introduced a quantity $\epsilon_g$ as a characterization of the violation of gauge symmetry, and used it as the key element to his analysis of the simulation cost.  

These issues were conveyed to Hank through FNAL, and later, also directly to him.
Here was (a part of) the scientific part of our message:

\begin{itemize}
\item
\textit{All terms from (1) to (5) are separately gauge-invariant. For a complex link variable $Z_{j,\vec{n}}$ connecting sites $\vec{n}$ and $\vec{n}+\hat{j}$, 
the gauge transformation is $Z_{j,\vec{n}}\to\Omega^{-1}_{\vec{n}}Z_{j,\vec{n}}\Omega_{\vec{n}+\hat{j}}$, where $\Omega_{\vec{n}}$ and $\Omega_{\vec{n}+\hat{j}}$ are special unitary matrices~\cite{Kaplan:2002wv,Bergner:2024qjl}. }

\item
\textit{The term (3) is not a Gauss-law-like constraint enforcing tracelessness. This is a part of the scalar kinetic term~\cite{Kaplan:2002wv}. It is easy to see it noting that $Z\bar{Z}$ is $W^2=\exp(2a\phi)$ up to a normalization factor. }

\item
\textit{
Therefore, a quantity $\epsilon_g$ defined in ref.~\cite{Lamm:2026zzl} has nothing to do with the violation of gauge symmetry --- gauge symmetry is exact at any lattice parameter. Rather, a departure of $\textrm{Tr}(W-\textbf{1})$ from zero (which is closely related to $\textrm{Tr}(W-\textbf{1})^2$ when the departure is large) is related to an effective shift of the lattice spacing. Note that we can add an additional term $-\gamma\mathrm{Tr}(Z\bar{Z})$ to the Hamiltonian so that $\textrm{Tr}(W-\textbf{1})$ can be tuned to zero without having a large mass~\cite{Halimeh:2024bth,Mendicelli:2026sib}.  }

\end{itemize}

In the second version, the elementary errors associated with the gauge symmetry were partially corrected based on our comments, as we can see from footnote 1 in ref.~\cite{Lamm:2026zzl}.
Still, however, a wrong interpretation of $\epsilon_g$ is adopted and used as the key element in sustaining the cost analysis in the first version of ref.~\cite {Lamm:2026zzl}. The key idea of Kaplan, Katz, and \"{U}nsal~\cite{Kaplan:2002wv} (which is a version of dimensional deconstruction proposed by Arkani-Hamed, Cohen, and Georgi~\cite{Arkani-Hamed:2001kyx}) was that the vacuum expectation value of the complex matrix generates a lattice dynamically. Therefore, a departure of $\textrm{Tr}(W-\textbf{1})$ from zero (which is closely related to $\epsilon_g=\textrm{Tr}(W-\textbf{1})^2$) merely describes an effective shift of the lattice spacing from a bare input value.\footnote{
Imagine $\mathrm{SU}(2)=\mathrm{S}^3$ is embedded into $\mathbb{R}^4$. Regardless of the value of the radial coordinate, the angular direction is always SU(2). 
}$^,$\footnote{See ref.~\cite{Bergner:2026duh} for details with examples from numerical simulations. 
} This fact falsifies the scaling argument presented in ref.~\cite{Lamm:2026zzl}. \\
\\
We emphasize that the ultimate goal of the entire community --- whether through orbifold lattice (Kaplan-Katz-\"{U}nsal) or Kogut-Susskind formulations --- remains the realization of precise and efficient quantum simulations of gauge theories. We hope this discussion serves as a catalyst for more robust and foundational cross-comparisons that will bring us closer to that shared objective.
\\
\\
We are happy to answer any questions and engage in a positive scientific discussion in separate communications.
\begin{center}
{\Large \textbf{Note added}}
\end{center}
Several colleagues asked whether ref.~\cite{Lamm:2026zzl} provides any insight, given that already the interpretation of eq.~(1) in that work does not apply to the present setting. We believe there is still a useful lesson.

Ref.~\cite{Lamm:2026zzl} states that no explicit circuit construction is available for the orbifold lattice. In reality, however, explicit quantum circuits for generic gauge groups, dimensions, and lattice geometries are known analytically, based solely on quantum Fourier transforms, CNOT gates, and single-qubit rotations~\cite{Halimeh:2024bth,Halimeh:2025ivn}. The only thing left is a simple, device-specific compilation. The circuit structure is universal: Yang-Mills theory and the harmonic oscillator share the same structure. This universality is precisely what makes it possible to perform cost estimates for any truncation level and any SU($N$).

The cost estimates for the Kogut–Susskind formulation in ref.~\cite{Lamm:2026zzl} highlight a different point. Very special cases --- such as SU(2) at the lowest truncation level, the large-$N$ limit under assumptions about low-energy states, or discrete subgroup constructions that do not reproduce the continuum limit --- are analyzed as ``competitors" to the continuum limit of generic SU($N$) realized by the orbifold lattice. This reflects the current difficulty of writing explicit quantum circuits for the continuum limit of generic SU($N$) --- including even SU(3) --- within the Kogut–Susskind framework~\cite{Hanada:2025yzx}.

In this sense, the comparison in ref.~\cite{Lamm:2026zzl} illustrates a structural limitation of the Kogut-Susskind approach at present.
\begin{center}
{\Large \textbf{Acknowledgment}}
\end{center}
We thank several colleagues for helpful discussions, while choosing not to participate in the April 1$^{\textrm{st}}$ tradition; Hank Lamm for clarifying his understanding of the subject via private communication; and the STFC for the support through the consolidated grant ST/Z001072/1. 
Finally, we thank FNAL for prompt and appropriate intervention that eliminated the unprofessional tone in the first version of ref.~\cite{Lamm:2026zzl}.

\bibliographystyle{utphys}
\bibliography{ref}

\end{document}